# Insulator to Metal Transition in $WO_3$ Induced by Electrolyte Gating

X. Leng[1], J. Pereiro[1,2], J. Strle[1,3], G. Dubuis[1,4], A. T. Bollinger[1], A. Gozar[5], J. Wu[1], N. Litombe[6], C. Panagopoulos[2], D. Pavuna[4] & I. Božović[1,5]

Tungsten oxide and its associated bronzes (compounds of tungsten oxide and an alkali metal) are well known for their interesting optical and electrical characteristics. We have modified the transport properties of thin $WO_3$ films by electrolyte gating using both ionic liquids and polymer electrolytes. We are able to tune the resistivity of the gated film by more than five orders of magnitude, and a clear insulator-to-metal transition is observed. To clarify the doping mechanism, we have performed a series of incisive *operando* experiments, ruling out both a purely electronic effect (charge accumulation near the interface) and oxygen-related mechanisms. We propose instead that hydrogen intercalation is responsible for doping $WO_3$ into a highly conductive ground state and provide evidence that it can be described as a dense polaronic gas.

[1]Brookhaven National Laboratory, Upton, NY 11973, USA. [2]Division of Physics and Applied Physics, School of Physical and Mathematical Sciences, Nanyang Technological University, 637371, Singapore. [3]Jožef Stefan Institute, Jamova 39, 1000 Ljubljana, Slovenia. [4]Ecole Polytechnique Fédérale de Lausanne, CH-1015 Lausanne, Switzerland. [5]Yale University, New Haven, CT 06520, USA. [6]Harvard University, Cambridge, MA 02138, USA. Correspondence should be addressed to I.B. (e-mail: bozovic@bnl.gov).

**INTRODUCTION**

Electrolyte gating has attracted much attention lately as an alternative to chemical doping to tune the carrier concentration in various materials.[1-23] It enables large changes in the sheet carrier density, as high as $10^{15}$ cm$^{-2}$, which can be utilized for detailed studies of electronic phase transitions and search for new superconducting phases. The mechanisms at play in this technique, however, are under debate. One view is that the charging process is purely electronic. In this case two electric double layers are formed, one at the gate and one at the material being gated, as the ions in the electrolyte are drawn to these two surfaces. The strong electric fields created at these two locations are screened on the scale of the screening length (~1 nm) by mobile charge carriers drawn into the gate and the material being studied, changing their carrier concentrations.[3-8,12,23] On the other hand, in recent experiments, it has been observed that entire gated films of oxides, 100 nm thick and even more, became metallic upon electrolyte gating.[10,11,14,16] This apparent paradox has led to the competing claim that the main mechanism of sample charging is in fact an electrochemical process, such as oxygen intercalation or the formation of oxygen vacancies induced by the strong electric field, as indeed thoroughly substantiated in the case of $VO_2$[11] and $SrRuO_3$.[16] Furthermore, some studies have also shown that this dichotomy in mechanisms does not always strictly hold and that both processes can be important in affecting changes in materials.[15,18,20]

Tungsten oxide ($WO_3$) is a natural candidate for electrolyte gating measurements as it has been widely studied for its special optical and electrical properties[24,25] and can be chemically doped by a large variety of elements, forming an extensive family of compounds known as tungsten bronzes. Some of these compounds are superconducting,[26-30] with the transition temperature ($T_c$) in the range from 0.5 K to 7 K. More alluring, Reich *et al*. have reported a two-dimensional (2D) superconducting phase at the surface of Na-doped WO$_3$, with $T_c \approx 90$ K.[31,32] Evidence for similar localized but disconnected superconducting regions with $T_c \approx 125$ K have also been reported in Na$_x$WO$_{3-y}$ infiltrated into nanoporous materials.[33] A few groups have recently performed electrolyte gating work with ionic liquids on WO$_3$,[19-22] varying the resistivity by at least an impressive 8 orders of magnitude and producing transitions from insulating to metallic behavior when the sample sheet resistance is around 1 kΩ/□. None, however, have observed any superconducting phases. Several explanations for how the excess conductivity is created in WO$_3$ by electrolyte gating have been put forth from electrostatic,[19,22] to a mix of electrostatic and electrochemical processes,[20] to a structural phase transition driven by oxygen removed from WO$_3$ by the applied electric field,[21] with no apparent consensus.

In the present study, we performed extensive charging experiments on atomically flat WO$_3$ films using either polymer electrolytes or ionic liquids as gate dielectrics. After the samples are fully charged, we observe large drops in the sheet resistance, and a clear insulator-to-metal transition. Although we have densely covered a large range of doping levels, we have seen no indications of superconductivity down to 300 mK. Our experiments on films of different thicknesses indicate that the entire films are affected by the electrolyte gating, and hence the charging process cannot be purely electrostatic. By further experiments we have demonstrated that oxygen electromigration is not involved in the charging process, either. Instead, we propose and provide evidence for charging dominated by hydrogen (specifically H$^+$) intercalation.

**RESULTS**

The WO$_3$ films were synthesized by rf-magnetron sputtering on single-crystal YAlO$_3$ (YAO) substrates (see Methods). The small lattice mismatch enabled us to synthesize high quality WO$_3$ films with atomically flat surfaces (Fig.1). The films were patterned in Hall-bar configuration devices (Fig. S1) of different sizes using microfabrication techniques. Two different electrolytes were used in the experiments. One was a polymer electrolyte, sodium fluoride salt dissolved in 1,000 molecular weight polyethylene glycol (PEG-NaF). It is a wax-like solid at room temperature and can be charged (*i.e.*, it acts as a conductive electrolyte) at $T = 305$ K. The other electrolyte was the ionic liquid DEME-TFSI.[5] It is liquid at room temperature but freezes to a glassy state at $T = 240$ K. In our experiments, we changed the gate voltage ($V_G$) at the charging temperature, kept it at that temperature for 30 minutes, and then slowly cooled down to liquid helium temperature, while measuring the device resistance.

In Fig. 2 we show the temperature dependence of the resistivity ($\rho$) at various gate voltages for four films of different thicknesses. We used the ionic liquid DEME-TFSI for sample C and the polymer electrolyte PEG-NaF for the other three samples. Despite the differences, the transport data for these four samples are very similar to each other. The original samples ($V_G = 0$ V) are very insulating, with the resistivity exceeding 10 Ω-cm at 250 K. When a positive gate voltage is applied, the resistance drops, indicating that mobile electrons have been induced in the sample. The highest gate voltage we used was in the range 2.0-2.2 V; at higher voltages, the resistance saturates. We usually keep the samples charged at the highest voltage for several hours until the resistance shows no further changes. At the highest gate voltage, the samples reach a metallic state with a positive temperature slope of resistance ($d\rho/dT > 0$). The sheet resistance drops more than 5 orders of magnitude at $T = 200$ K, and much more at lower temperatures, comparable to the change in resistivity observed in previous electrolyte gating studies on $WO_3$.

The resistivity of a charged and saturated $WO_3$ film at $T = 200$ K is about 110 μΩ-cm, which is three times smaller (*i.e.*, more metallic) than what we see in optimally doped cuprate films showing high-$T_c$ superconductivity. However, we did not see any signs of superconductivity, down to $T$ = 300 mK, in any of the WO$_3$ films we have studied so far (Fig. S2). Given that we have covered a wide range of doping levels, this seems to contradict the claims of 2D high-$T_c$ superconductivity at the surface of Na-doped WO$_3$.[31,32] A (remote) possibility is that doping by Na may not only provide mobile carriers but also cause unidentified structural changes, presumably crucial for superconductivity yet absent in the case of electrolyte gating. Reproducible bulk (3D) superconductivity has been observed in Na-doped WO$_3$ but only at very low temperatures, with the maximum $T_c$ = 0.57 K.[26] However, we did not see this low-temperature superconductivity either.

A clear metal-insulator transition (MIT) is seen in the electrical transport data taken on each of our samples. The insulating ($d\rho/dT < 0$) and metallic ($d\rho/dT > 0$) regimes are separated by the critical sheet resistance $R_C \approx 200\ \Omega/\square$. Within the purely electronic charging scenario, the electrolyte gating should only affect the sample surface, and hence one would expect a 2-dimensional (2D) MIT, for which $R_C$ should be close to the universal resistance quantum $R_Q = h/e^2 = 25.8\ k\Omega/\square$.[34] However, the critical sheet resistance we observe here is two orders of magnitude smaller, suggesting that the whole film might have been affected.

We fabricated and studied a step-like film, as illustrated in the inset of Fig. 3, in order to determine the depth over which a WO$_3$ sample can be affected by electrolyte gating. We have charged the whole film with PEG-NaF to the saturating point, then measured the conductance of each section simultaneously. In Fig. 3, we can see that the saturated sheet conductance $G_S$ increases almost linearly with the film thickness, $d$. Therefore, the conductivity $\sigma = G_S/d$ is essentially constant for sections of different thicknesses, which suggests that the whole sample has been affected by the electrolyte gating, even for the thickest (70 nm) section. All other measurements performed on films with uniform thicknesses (Fig. 2) are in qualitative agreement with the results in Fig. 3. Given that the Thomas-Fermi screening length in a metal is very short (< 1 nm), the electric field cannot penetrate through a 70 nm thick metallic film, and hence this finding rules out a purely electrostatic effect. Some chemical process that affects the entire film must be involved.

To check whether oxygen electromigration is playing a role in our experiments, we performed a series of charging/discharging experiments in vacuum ($10^{-6}$ Torr overall, oxygen partial pressure less than $10^{-7}$ Torr) and in air (oxygen partial pressure close to 155 Torr). We cycled the gate voltage with time, occasionally switching the environment between vacuum and air. The charging temperature was fixed at $T = 305$ K and the sample resistance was measured at that temperature during charging/discharging processes. The results are shown in Fig. 4. The top panel shows the changes in the gate voltage, and the bottom panel shows the concomitant changes in the resistance. One can see that both the charging and discharging processes are reversible and reproducible. However, there is no apparent difference between the charging-discharging cycles performed in vacuum and in air, in contrast to what was observed in $VO_2$[11] and $SrRuO_3$[16] experiments, as well as one report with $WO_3$.[21] This implies that oxygen electromigration is not relevant when discussing the resistivity changes observed in our electrolyte gating experiments on $WO_3$.

The $WO_3$ structure can be viewed as an $ABO_3$ perovskite where tungsten occupies the B site but the A site remains empty. Thus, it is relatively easy to intercalate small atoms (hydrogen or alkali metals) to occupy the A site as was demonstrated in much of the early work on $WO_3$.[35] Since we are using PEG-NaF as electrolyte, intercalation of $Na^+$ is conceivable. To check this possibility, we have tried gating with pure polyethylene glycol (adding no salts whatsoever) on WO$_3$; the outcome turned out to be the same (see Fig. S3), ruling out sodium intercalation. This leaves us with the possibility that the primary effect of electrolyte gating is hydrogen (H$^+$) intercalation, as it is well known that hydrogen intercalation, i.e. protonation, can make WO$_3$ metallic.[35] While we made every best effort to dehydrate our electrolytes under vacuum before applying them to our samples and to minimize the time between removing the electrolyte from vacuum, applying it to our samples, and beginning the transport measurements, we believe that the source of hydrogen ions is the self-ionization of water that the electrolytes absorbed in the short time that they were exposed to air. Indeed, several studies have used hygroscopic solids as a proton source to controllably modify the conductivity of WO$_3$ in a transistor geometry[36-38] so this means of conductivity enhancement in WO$_3$ that is covered with a hygroscopic electrolyte cannot be overlooked. We note that controlled protonation can be used to tune the conductivity of other oxides as well, as has recently been demonstrated in VO$_2$,[39,40] and In$_2$O$_3$/ZnO.[41]

We performed a comparison of *operando* XRD patterns measured during electrolyte charging/discharging processes with XRD data taken after vacuum, ozone and hydrogen annealing procedures, shown in Fig. 5, to directly compare to samples where hydrogen was added in the more traditional way and also to confirm that oxygen vacancies are not at play here. The *operando* XRD curves measured during the charging and discharging processes (Fig. 5a) show negligible shift (~0.01°) of WO$_3$ Bragg peaks; the change of the lattice constant is smaller than 0.05%. This differs with previous XRD data taken while electrolyte gating WO$_3$ that showed the lattice constants changing by at least 2%.[21,22] However, we point out that there is really no agreement on this particular point as the *c*-axis lattice constants have been observed to both decrease[21] and increase[22] while charging in the range of gate voltages used here (up to +3 V). Several differences between our work and that of Refs. 21 and 22 exist that could account for this disparity in the XRD data. Different combinations of substrates and growth temperatures were used in each of these three studies, which affects the resulting film structure, quality, and epitaxial strain. Also, different electrolytes and measurement environments were used which could contribute to the differences in the XRD results. Regardless, we clearly find that the $WO_3$ unit cell in our work stays essentially the same during the charging process, even though the sample has been driven to the metallic state (Fig. S4). If there were any cations intercalated/doped into the film during this process, they must be small enough to maintain the original $WO_3$ lattice constants.

In contrast, annealing in vacuum at 550 °C shifts the $WO_3$ Bragg peak to a lower angle by 0.16°, corresponding to a 0.7% change in the lattice constant (the top panel of Fig. 5b), which means that the film structure has changed significantly. Vacuum annealing at a lower temperature ($T \leq 520$ °C) did not change the XRD pattern at all. Moreover, the sample modified by high-temperature annealing in vacuum can be brought back towards the original state by annealing in ozone at 450 °C for just 1 hour (the bottom panel of Fig. 5b). Both the Bragg peak and the thickness fringes are moving back towards the XRD curve measured for the as-grown film. This points to changes in oxygen stoichiometry, and shows that we can clearly detect them if they are present, and that they do not occur in our films except at high temperatures.

In Fig. 5c, we present XRD curves measured for a sample annealed in hydrogen. We have tuned the annealing temperature from 300 °C to 520 °C, and yet the $WO_3$ Bragg peak did not shift at all, although the fringes smear out after annealing at the highest temperature, indicating some surface degradation. The XRD patterns in Fig. 5a (charging) and 5c ($H_2$ annealing) are similar to each other while the ones in Fig. 5b (vacuum annealing) are very different. This indicates that hydrogen has been involved in the charging process, comparable to the hydrogen annealing process. Furthermore, we have also measured the transport properties of the WO$_3$ film after the hydrogen annealing procedures as illustrated in Fig. 2 by empty squares ("□", H$_2$ annealed at 300 °C) and empty circles ("○", H$_2$ annealed at 520 °C). When the resistivity reaches a certain level, the $R(T)$ curves of the electrolyte-gated and hydrogen-annealed samples show almost the same slope, which suggests the changes in the sample induced by the electrolyte gating are the same as those induced by the hydrogen annealing.

We performed one additional transport experiment in which we covered a portion of a WO$_3$ film patterned into a Hall bar with an electrolyte and looked for possible changes in the resistivity nearby in areas of the bar that were not covered by the electrolyte. If a change were to be seen, this would clearly point to electromigration; given the small probe voltage and the relatively large distances involved, the longitudinal (in-plane) electric field is negligible compared to the minimum needed for an observable electronic field effect. Since this portion of the sample stays essentially charge neutral, any increase in the mobile electron density must be exactly compensated by intercalation of positive ions. In Fig. S5, we show that in a region of the Hall bar that is 300 µm away from the electrolyte the resistance drops by 70%. This experiment clearly demonstrates that doping takes place by cation intercalation. Note that this effect is seen only if the applied gate voltage is positive, *i.e.*, if the electromigrating ions are positively charged. We know that they must also be very small based on our x-ray experiments, which uniquely points to hydrogen.

To evaluate how many hydrogen atoms are intercalated into the WO$_3$ films, we need to measure the carrier density of the charged and saturated sample. For ordinary semiconductors, the standard method is to measure the Hall effect and infer $n$ from the Hall number, $R_H = 1/ne$, where $e$ is the electron charge. However, the Hall effect measurement of our WO$_3$ films gives too high a value if we assume one type of carrier, $n = 2.8$ electron per one $WO_3$ formula unit (see Fig. S6), which is questionable on physical grounds. (Similarly large carrier densities were also determined from Hall effects measurements in Ref. 22.) One possible explanation for the low Hall number from which the high carrier density was derived is that more than one type of charge carrier is present. Another is that this may point to polaron formation, since it is known that electron-phonon coupling is very strong in many transition metal oxides, including $WO_3$ specifically[42], and that for polaronic materials the above simple Hall formula is not applicable.

We modified a Fourier transform infrared (FTIR) spectrometer to allow *operando* measurements during the charging/discharging processes to test the polaronic scenario. Typical absorption spectra obtained from a charging process are shown in Fig. 6a. During the charging process a strong IR band appears around 0.45 eV (see Supplementary for fitting details). In these experiments DEME-TFSI was used as the electrolyte and the cut-off observed at 0.2 eV is due to a glass cover slip that we used to hold the ionic liquid in place. We have also used Teflon foil instead of the glass and we can see a similar peak around 0.45 eV (Fig. S7). This type of feature has been interpreted as a telltale signature of polaron formation in a large number of oxide materials and their energy was assigned to the polaron binding energy.[43,44] Hence, our IR data support the idea that hydrogenation leads to the formation of a polaronic gas. Furthermore, while similar mid-IR peaks were observed in both absorption[44] and photo-induced measurements[45] in $WO_{3-\delta}$ samples, the peak energy in Fig. 6 is in better agreement with the latter experiments. A natural explanation is that in hydrogenated films the absorption would correspond to a Frank-Condon transition from a (reduced) $W^{5+}$ ion (like in photo-doping) rather than a transition of an electron trapped at a vacancy site in oxygen deficient samples.

As the gate voltage is applied, the spectral weight of the IR polaronic band increases monotonically with time and finally saturates. The 3D carrier concentration at saturation extracted from the plasma frequency associated with the IR band (Fig. 6b) renders a value $n \sim 4.4 \times 10^{21}$ cm$^{-3}$, which corresponds to approximately 0.2 carriers per WO$_3$ primitive unit cell. It is conceivable that we have reached the maximum for this mechanism; if only the nearest neighbor W-O lattice distortions were involved, the absolute upper bound for non-overlapping small polarons would be 0.5. Within our scenario, if an even higher doping could be achieved, it could trigger a collapse of small polarons and a transition into a state of matter not explored before.

Although the sample has been charged to the metallic state, the nominal Hall mobility is quite low, $\mu_H = R_H/\rho \approx 1$ cm$^2$/Vs. Such a low mobility is typical of small polarons, and in fact it has been reported[42] that polarons of small or intermediate size form in some structural phases of WO$_3$. Since hydrogen intercalation induced by the electrolyte gating process is slow, reversible, and reproducible for all of the properties we have measured so far, including transport (Fig. 4), XRD (Fig. 5) and IR absorption (Fig. S8), electrolyte-gated WO$_3$ may prove to be a convenient platform to study polaron physics, including the elusive polaronic metal state.

**DISCUSSION**

While we believe that we have presented a strong case that hydrogen is involved in the electrolyte gating of WO$_3$ films, the mechanism of electrolyte gating might well be different in other materials. A case in point is VO$_2$ or SrRuO$_3$, as previously mentioned. Both are very sensitive to oxygen; annealing in vacuum or oxygen significantly changes their oxygen stoichiometry, crystal structure, and carrier densities. Electric field induced electromigration of oxygen will therefore clearly play a role in electrolyte gated VO$_2$ or SrRuO$_3$ devices and may also be relevant when gating other oxygen sensitive materials such as SrTiO$_3$, InO$_x$, or YBaCuO$_{7-x}$. In contrast, WO$_3$ seems to be relatively stable against the formation of oxygen vacancies as we find that annealing in vacuum at 520 °C for 2 hours does not change the oxygen stoichiometry at all. Still, because of its vacant center (A) site, WO$_3$ is very susceptible to intercalation of hydrogen and small alkali cations, which must be considered when gating this material. The larger narrative of this and much of the recent electrolyte gating work seems to be that no one mechanism – purely electrostatic, oxygen vacancies, lattice deformations, intercalations, etc. – is universal in all situations. Even thin film samples of the exact same material can be affected by electrolyte gating in different ways depending on how it was prepared (*e.g.* epitaxial or polycrystalline films, based on lattice match of substrates) as has been demonstrated in SmNiO$_3$[12,13] and VO$_2$.[46] All of these possibilities must evidently be considered and carefully investigated before making any broad conclusions about the mechanisms at work in any electrolyte gated materials.

## METHODS

**Thin film synthesis.** Atomically flat WO$_3$ films have been synthesized by r.f. magnetron sputtering. The growth temperature was 750–850 °C and the pressure during growth was about 60 mTorr with an O$_2$/Ar ratio of 4:1. The thickness was determined using X-ray reflectivity data, and the growth rate was about 1 nm/minute. The substrates we used were YAlO$_3$ (YAO) single crystals with the surfaces polished perpendicular to the crystallographic [110] direction. YAO has an orthorhombic structure with lattice constants $a_0$ = 5.176 Å, $b_0$ = 5.307 Å, and $c_0$ = 7.355 Å. Along the [110] direction of YAO, the lattice mismatch between YAO and the γ-monoclinic WO$_3$ is +0.7% (7.355 Å *vs.* 7.306 Å) along the *a*-axis and –1.7% (7.413 Å *vs.* 7.540 Å) along the

*b*-axis. The small lattice mismatch enabled us to synthesize high quality $WO_3$/YAO films with atomically flat surfaces, as confirmed by the XRD and AFM results.

**Transport measurements.** The $WO_3$ films were patterned into a set of Hall bar devices of various aspect ratios with gold contacts. We used photolithography for larger features and electron-beam lithography for features smaller than 1 µm. The gate, a piece of thin platinum mesh, was situated above the Hall bars for the transport measurements (Fig. S1a) with both ionic liquid DEME-TFSI and polymer PEG-NaF used as electrolytes. Note that resistance was measured using delta method averaging to eliminate the effects of offset currents, such as gate leakage.

**Optical measurements.** For *operando* XRD and FTIR measurements we used side-gated devices (Fig. S1b). Gold contacts were sputtered at the four corners of the sample in a van der Pauw configuration. These four probes were connected and grounded at the beginning of the electrolyte gating to make sure they were at the same equipotential surface. After the sample was charged to a low-resistivity state these four contacts were used to measure the changes in resistivity during further charging. We used polymer electrolyte PEG-NaF for XRD measurements, and the ionic liquid DEME-TFSI for FTIR measurements. A glass coverslip or Teflon foil was used to hold the ionic liquid during the FTIR measurements.

**DATA AVAILABILITY STATEMENT**

The data that support the findings of this study are available from the corresponding author upon reasonable request.

**ACKNOWLEDGEMENTS**

The experimental work reported here was performed at BNL and supported by U. S. Department of Energy, Basic Energy Sciences, Materials Sciences and Engineering Division. This research used resources of the Center for Functional Nanomaterials, which is a U.S. DOE Office of Science Facility, at Brookhaven National Laboratory under Contract No. DE-SC0012704. X.L. was supported by the Center for Emergent Superconductivity, an Energy Frontier Research Center funded by the U. S. DOE, Office of Basic Energy Sciences. G.D. and D.P. were supported by the Swiss National Science Foundation. J.P. and C.P. acknowledge financial support from the National Research Foundation, Singapore through Grant NRF-CRP4-2008-04. A.G. is supported by the Gordon and Betty Moore Foundation's EPiQS Initiative through Grant GBMF4410.

**COMPETING INTERESTS**

The authors declare no conflict of interest

**AUTHOR CONTRIBUTIONS**

X.L., J.P., and J.S. grew the $WO_3$ thin films. A.T.B and N.L. did the lithography. X.L. performed the transport measurements with help from A.T.B. and G.D. X.L., J.W., and A.T.B. performed XRD measurements. X.L. and A.G. performed the FTIR measurements. C.P. and D.P. advised on various parts of the project. I.B. conceived and supervised the project. X.L., A.T.B, and I.B. wrote the manuscript with feedback from all of the coauthors.

**FUNDING**

The experimental work reported here was performed at BNL and supported by U. S. Department of Energy, Basic Energy Sciences, Materials Sciences and Engineering Division. This research used resources of the Center for Functional Nanomaterials, which is a U.S. DOE Office of Science Facility, at Brookhaven National Laboratory under Contract No. DE-SC0012704. X.L. was supported by the Center for Emergent Superconductivity, an Energy Frontier Research Center funded by the U. S. DOE, Office of Basic Energy Sciences. G.D. and D.P. were supported by the Swiss National Science Foundation. J.P. and C.P. acknowledge financial support from the National Research Foundation, Singapore through Grant NRF-CRP4-2008-04. A.G. is supported by the Gordon and Betty Moore Foundation's EPiQS Initiative through Grant GBMF4410.

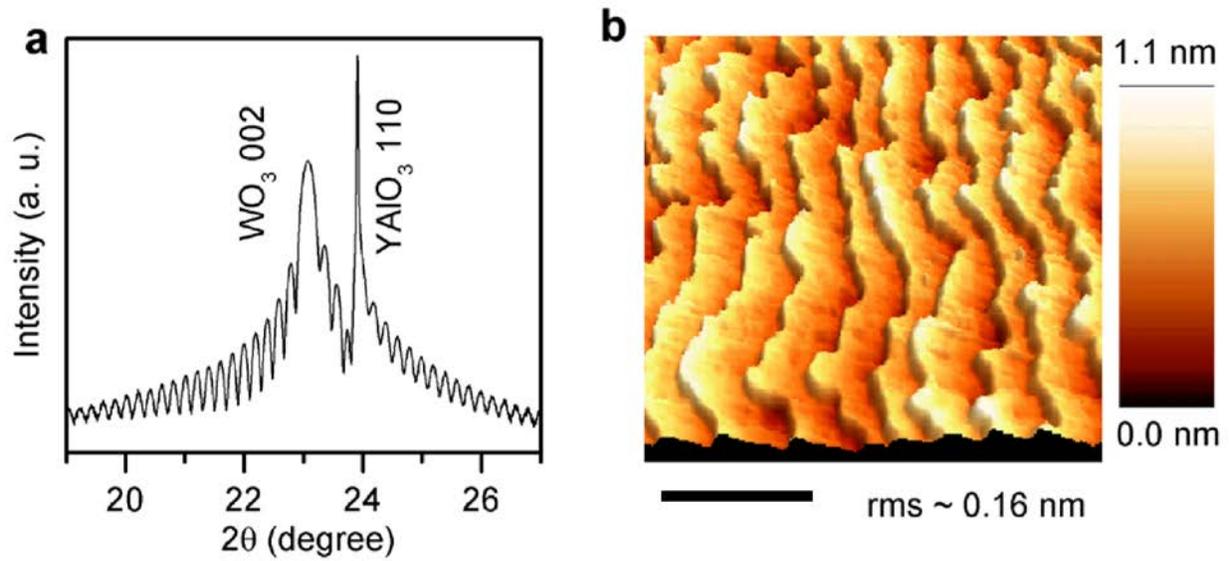

**Figure 1 | Structural characterizations of $WO_3$ films**. **a**, High-resolution Cu K$\alpha$ $\theta - 2\theta$ X-ray diffraction (XRD) pattern of a 45 nm thick $WO_3$ film deposited on a $YAlO_3$ (110) substrate. **b,** 3D atomic force microscope (AFM) image of a 14 nm thick $WO_3$ film deposited on a $YAlO_3$ substrate. Scale bar is 200 nm. The root mean square (rms) roughness is only 0.16 nm and the step-like terraces indicate the film has an atomically flat surface.

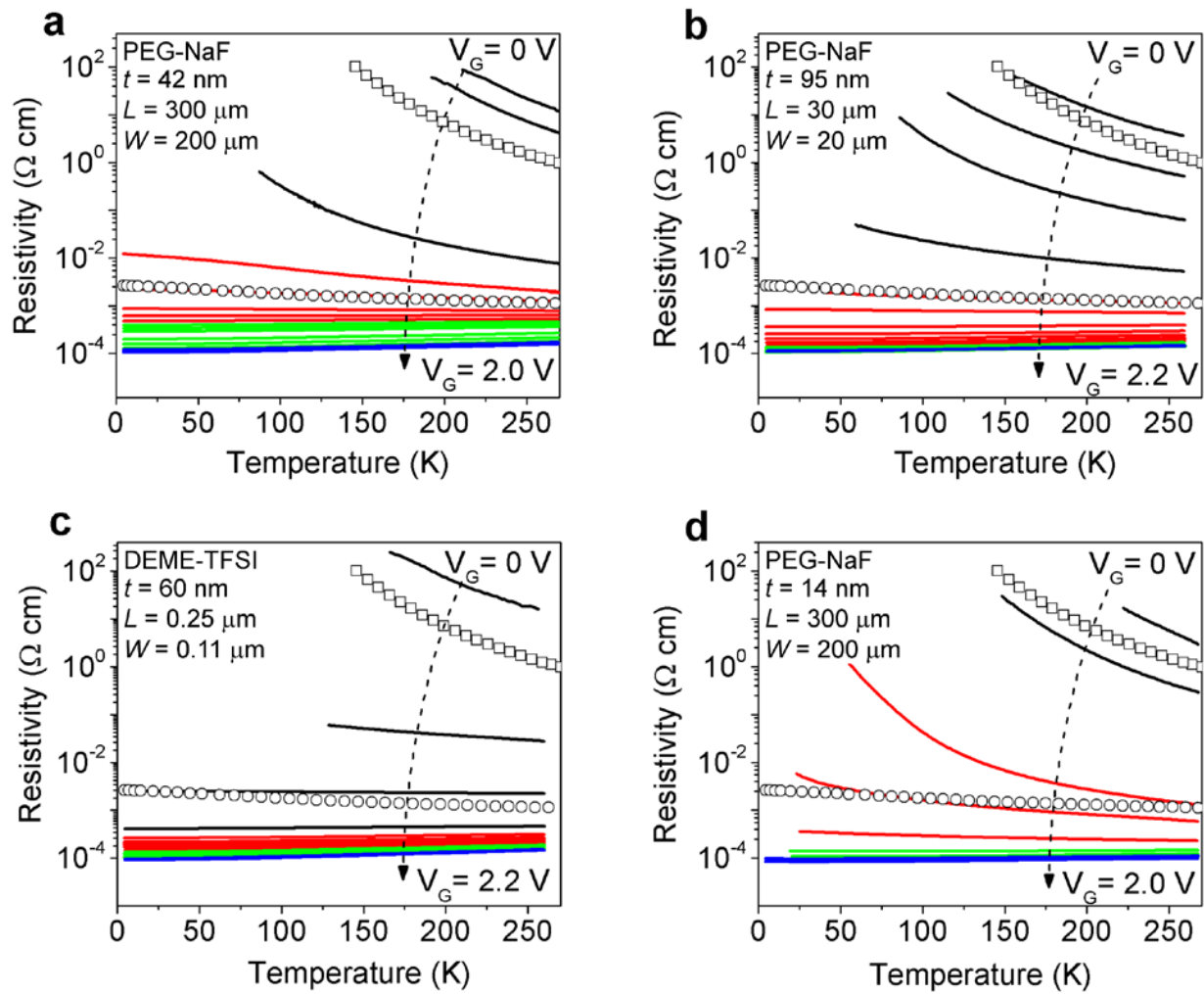

**Figure 2 | Resistivity vs. temperature of WO$_3$ samples as a function of gate voltage.** These four samples have different thicknesses (*t*), lengths (*L*) and widths (*W*): **a**, *t* = 42 nm, *L* = 300 µm, *W* = 200 µm; **b**, *t* = 95 nm, *L* = 30 µm, *W* = 20 µm; **c**, *t* = 60 nm, *L* = 0.25 µm, *W* = 0.11 µm; **d**, *t* = 14 nm, *L* = 300 µm, *W* = 200 µm. The gate voltage ($V_G$) was varied in steps of 0.1 V for sample A, B and D at 305 K and 0.2 V for sample C at 220 K. The electrolyte for sample C is DEME-TFSI while for the other three, it is PEG-NaF. As $V_G$ increases, the resistivity drops about 5 orders of magnitude at 250 K, and all four samples show a similar insulator-to-metal transition. Curves denoted by "□" and "○" are resistivity vs. temperature curves measured for a

30 nm thick WO$_3$ sample annealed in hydrogen forming gas at 500 °C and 520 °C for 1 hour in sequence.

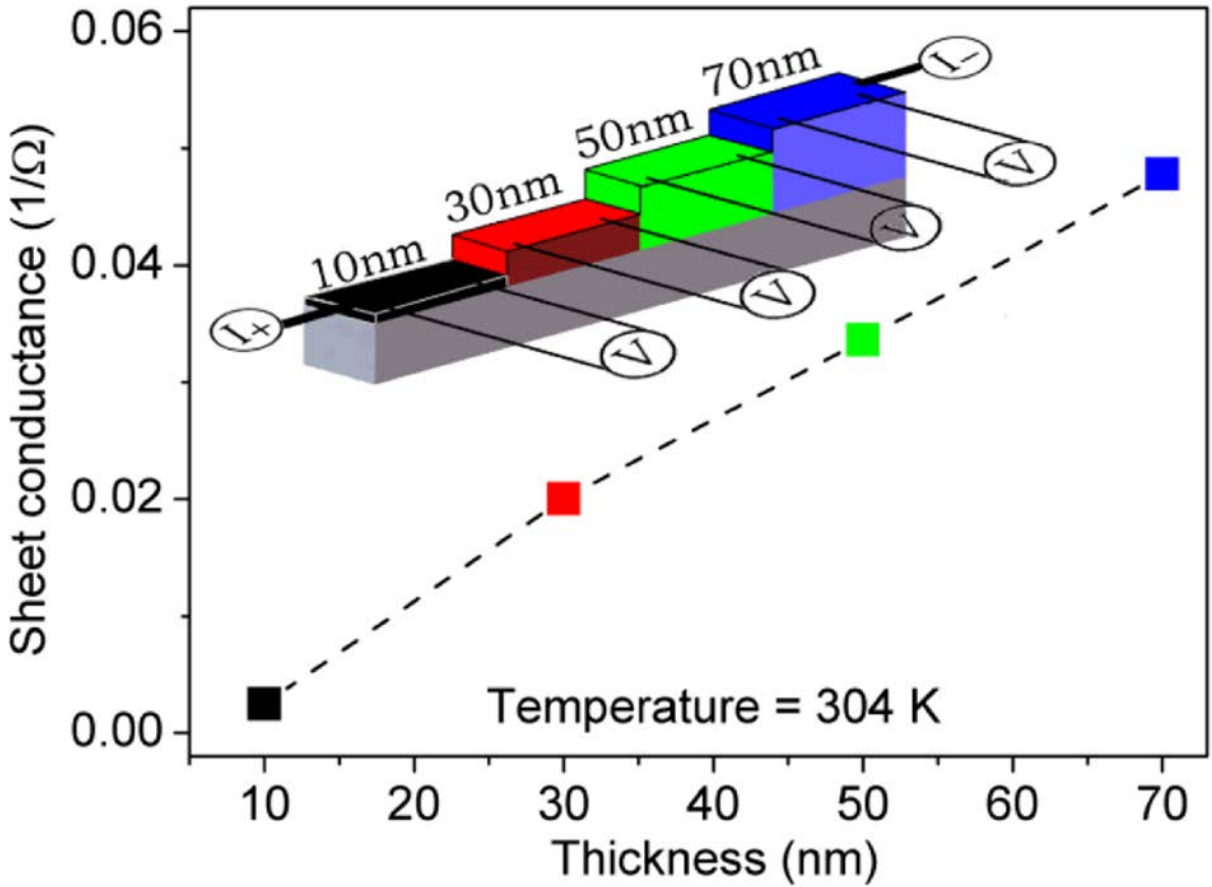

**Figure 3 | Maximum sheet conductance as a function of film thickness**. The electrolyte is PEG-NaF and the film has been charged at $V_G$ = 2.2 V and $T$ = 304 K for 2 hours. (Inset) Sketch of a step-like film, deposited using a shadow-mask that is moved in such a way that different portions of the films have different thicknesses.

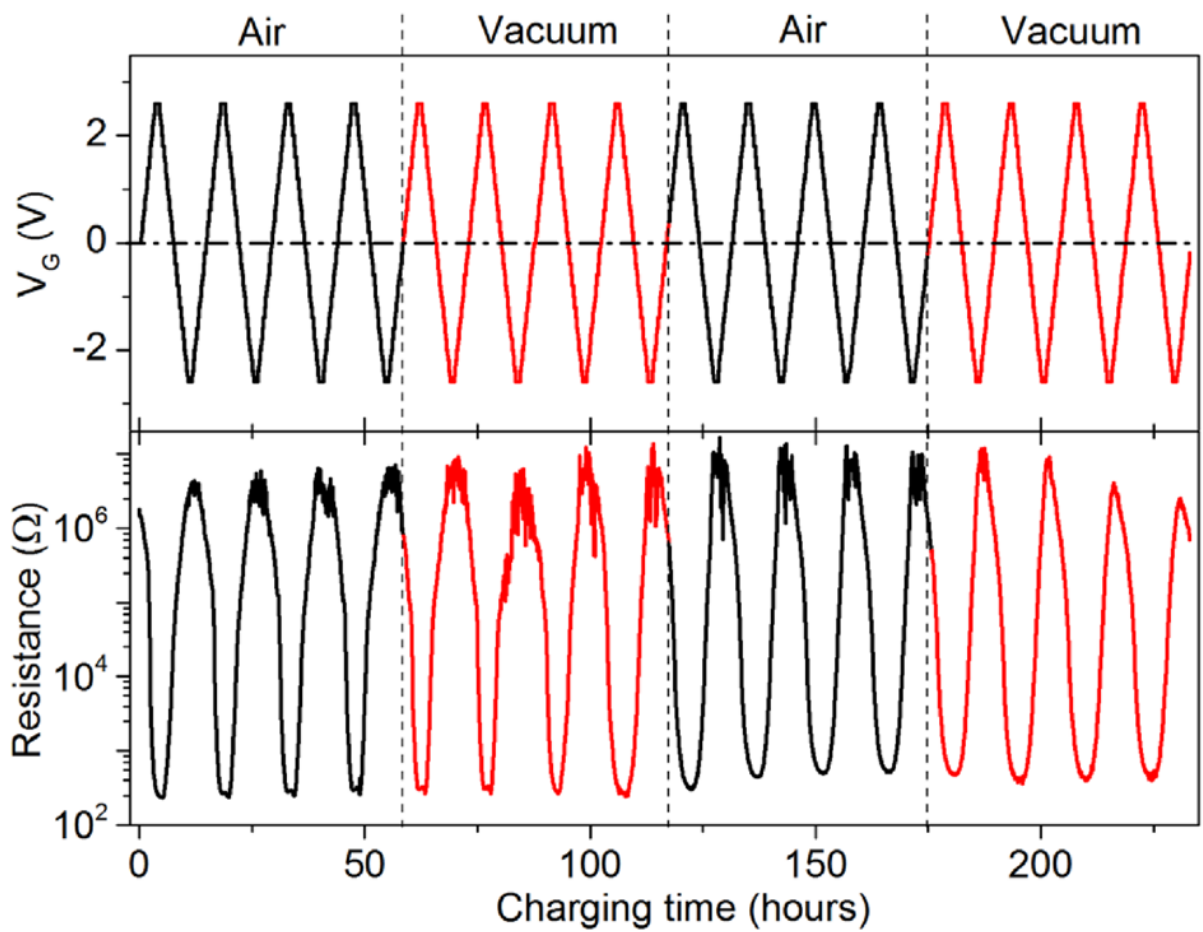

**Figure 4 | Sample resistance measured as a function of charging time as the sample is charged and discharged in air and in vacuum**. Top, the changes of the gate voltage ($V_G$) as a function of charging time. The electrolyte is PEG-NaF and the charging temperature is $T = 304$ K. The voltage was changed in steps of 0.2 V and the charging time for each step is 15 minutes. When $V_G$ increases from -2 V to 2 V, the sample is charged and when $V_G$ drops from 2V to -2 V the sample is discharged. For the black curves, the charging/discharging process occurs in air and for the red curves, the sample was charged/discharged in a vacuum of $10^{-6}$ Torr. Bottom, sample resistance measured as $V_G$ was cycled. The charging and discharging processes are very reversible and reproducible, and it apparently makes no difference whether the environment is air or vacuum.

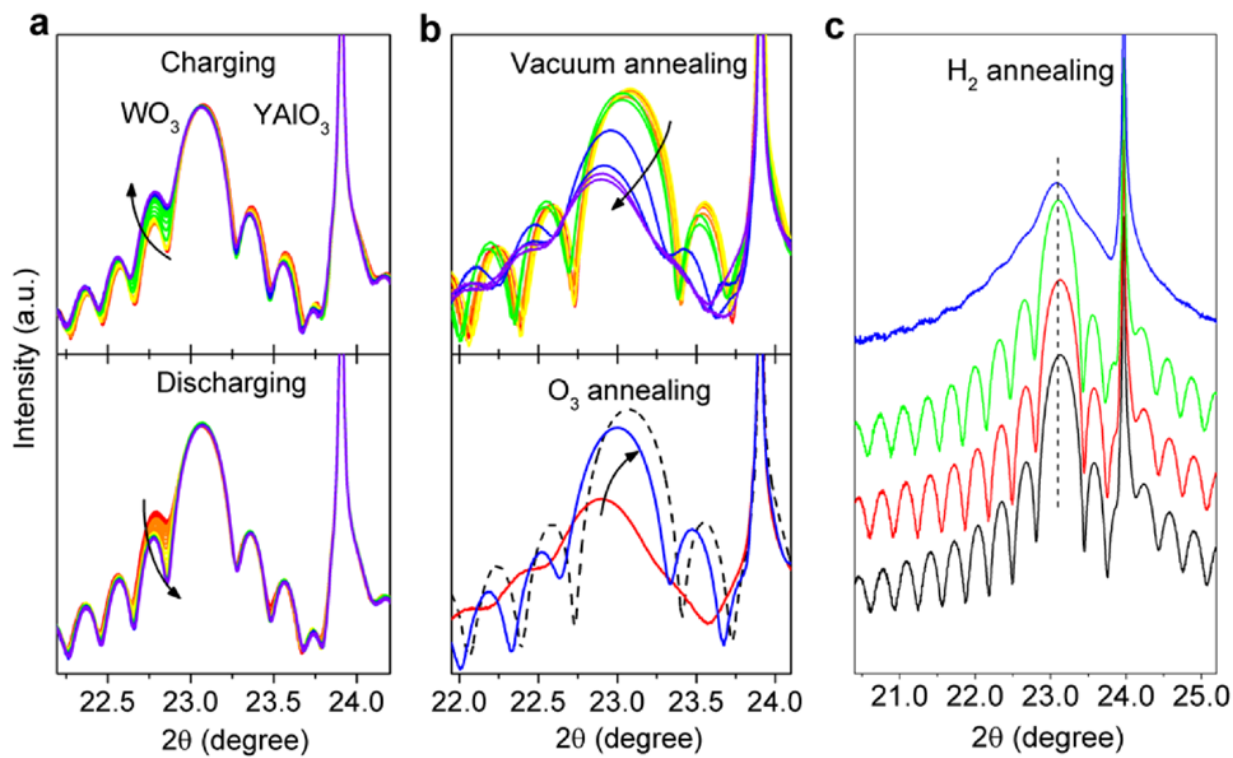

**Figure 5 | XRD curves measured during charging and discharging processes and compared with those measured after a series of vacuum, ozone, and hydrogen annealing procedures**. **a,** XRD curves measured during charging (top) and discharging (bottom) processes. The electrolyte is PEG-NaF and the curves are measured at room temperature. The maximum charging gate voltage is $V_G = 3$ V (-3 V for the discharging process). The (002) $WO_3$ Bragg peak shows negligible shifts (~0.01°) while the first fringe changes significantly. **b,** XRD curves measured after a series of vacuum and ozone annealing procedures in sequence. Top, following the direction of the arrow: the first one is the original (as-grown) film; the next 7 curves were measured after the film was annealed for 2 hours in $10^{-6}$ Torr vacuum at 400 °C, 400 °C, 450 °C, 450 °C, 500 °C, 500 °C and 520 °C; the last 4 curves were measured after the film was annealed at 550 °C for two hours each. No obvious changes can be seen if the annealing temperature is below 520 °C. However, increasing the annealing temperature to 550 °C shifts the (002) $WO_3$
25

Bragg peak to a lower angle by 0.16° and the thickness fringes smear out during this process. Bottom, the same $WO_3$/YAO film was then annealed in an $O_3$/$O_2$ mixture at 450 °C for 1 hour following the vacuum annealing procedure. The red curve is the one measured after the vacuum annealing procedure and the dashed line represents the original (as-grown) film. The blue curve is measured after ozone annealing and shows that the (002) $WO_3$ Bragg peak moves back towards the original position. (**c**) XRD curves measured after the film was annealed in hydrogen forming gas at different temperatures in sequence. From bottom to top: the first one is the original film, the next three are measured after the sample was annealed in hydrogen forming gas (mixture of $H_2$ and Ar) at 300 °C, 500 °C and 520 °C for 1 hour. The (002) $WO_3$ Bragg peak does not change at all during the hydrogen annealing process while the fringes smear out after the final annealing procedure.

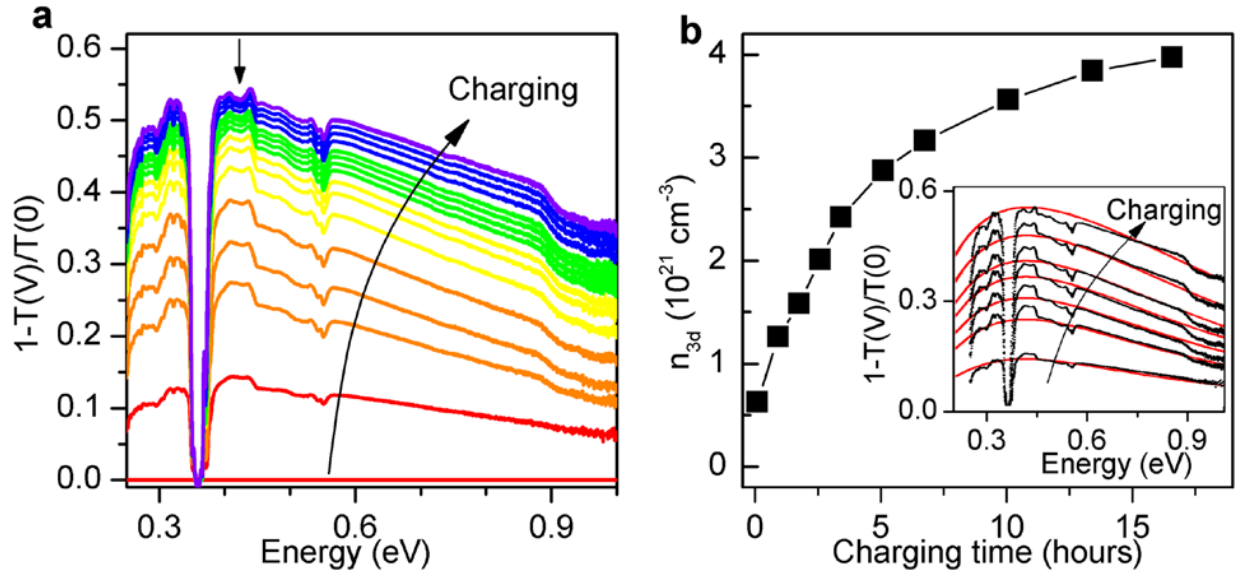

**Figure 6 | Fourier transform infrared spectroscopy (FTIR) data and fitting results**. **a,** Infrared absorption spectrum obtained from FTIR measurement during the charging process of a 45 nm $WO_3$ film. The electrolyte is DEME-TFSI and the charging was done at room temperature (294 K). At this temperature, 1.5 V is a gate voltage that we can safely apply according to our previous experience. The time interval between each curve is about 1 hour. The cut-off at 0.2 eV is due to a glass window that we used to hold the ionic liquid. If no gate voltage is applied, the total drift of the absorption spectra in 20 hours is below 0.05, *i.e.*, 10 times smaller than the change due to electrolyte gating. The short vertical arrow indicates the position of the broad IR peaks (taking up the entire spectral range shown) that corresponds to the polaron binding energy. **b,** 3D carrier concentrations as a function of charging time calculated from the fitting results. (Inset) Solid lines are the fitting results using the Lorentz oscillator model (see Supplementary).

## Supplementary Text

### Low temperature measurements
The transport measurements in the temperature range of 0.3 K to 40 K were performed in a $^3$He cryostat with a magnetic field up to 9 T. The film was first charged using a polymer electrolyte PEG-NaF at 305 K until the resistivity saturated, which usually takes several hours. Then it was cooled down to the lowest temperature (0.3 K) with the gate voltage on.

### Electrolyte gating
The maximum gate voltage one can apply on an electrolyte-gating device without inducing irreversible changes in the sample, also known as "the charging window", is the key parameter for achieving a reversible and repeatable gating process. Generally, the maximum gate voltage decreases as the charging temperature increases. In the transport measurements, the ionic liquid DEME-TFSI was charged at 220 K and the maximum gate voltage we could apply was 2.0–2.2 V. For the polymer electrolyte PEG-NaF, the charging temperature was 305 K and the maximum gate voltage falls into the same range.

The *operando* optical measurements (XRD and FTIR) were performed at room temperature. At this temperature, the maximum gate voltage for DEME-TFSI drops to 1.6 V while for PEG-NaF, it increases to 3.0 V. We used DEME-TFSI for the FTIR measurements and PEG-NaF for the XRD measurements and the gate voltage never exceeds their maximum values.

Another key parameter is the charging time. At a lower charging temperature or a lower gate voltage, the time needed to charge the sample to the saturation point becomes significantly longer. Thus in the XRD (PEG-NaF, charging at a lower temperature) and FTIR (DEME-TFSI, charging at a lower gate voltage) measurements, the charging time was several times longer than what we used in the transport measurements.

In the XRD measurements, each wide-angle XRD curve was measured in 2 hours. A total of 20 curves were taken during the charging process as the gating voltage increases from 0 to 3.0 V in a step of 0.1 V or 0.2 V. The gate voltage was then set to zero and the discharging process started immediately. Another 20 curves were taken during the discharging process with the gate voltage ramped from 0 to -3.0 V in a step of -0.1 V or -0.2 V. In the FTIR measurements, a total of 16 curves were measured and presented in Fig. 6a and the time interval between each curve is about 1 hour.

### Electromigration
The resistance of sections several hundred micrometers away from the charging area has been measured using the configuration presented in Fig. S5. In this configuration, the measuring current is just the gate current, which is several tens of nA. This is a DC mode measurement since the direction of the gate current never changes, which keeps the electromigration moving in the same direction throughout the measurement. However, this is different from what we used for the general transport (resistance and Hall) measurements. The transport measurements were carried out in a delta-mode to eliminate offsets thus the effect of electromigration is negligible. It is important to note that these results cannot be explained by a spread of the electrolyte over the neighboring area of the film. PEG-NaF is solid at room temperature and we thoroughly checked that the surface between the voltage pads in Fig. S5 remained pristine. The limited mobility of the protons also explains why the data in samples with uniform thickness, Fig. 2, are in full quantitative agreement with the results in the variable thickness film, Fig. 3. The voltage pads in Fig. 3 are much further apart than the diffusion length of the protons so "charge leakage" from neighboring regions with different thickness is negligible.

**FTIR data fitting**
The device shown in Fig. S1b has a stratified planar structure with 5 layers (4 interfaces) from top to bottom: air, ionic liquid, $WO_3$, $YAlO_3$, and air. The overall transmission coefficients can be calculated using a general formula that uses a complex generalization of the so-called elementary symmetric functions of the mathematical theory of polynomials:[1,2]

$$\mathcal{T}_4 = \frac{T_1 T_2 T_3 T_4}{1 + \bar{R}_1 R_2 + \bar{R}_1 R_3 + \bar{R}_1 R_4 + \bar{R}_2 R_3 + \bar{R}_2 R_4 + \bar{R}_3 R_4 + \bar{R}_1 R_2 \bar{R}_3 R_4}$$

Here, the complex transmission and reflection coefficients $T_j$ and $R_j$ are defined as:
$$T_j \equiv T_{j,j+1} = t_{j,j+1} \exp[-i\beta_j]$$
$$R_j \equiv R_{j,j+1} = r_{j,j+1} \exp[-2i(\beta_1 + \beta_2 + \cdots + \beta_j)]$$

The bar over $\bar{R}$ denotes changing all $\beta_j$ to $-\beta_j$.

Next, $t_{j,j+1}$ and $r_{j,j+1}$ are the Fresnel transmission and reflection coefficients of the $(j, j+1)$ interface:
$$t_{ij} = \frac{2n_i}{n_i + n_j}; \; r_{ij} = \frac{n_i - n_j}{n_i + n_j}$$

$\beta_j$ is the phase shift of the electromagnetic wave propagating through the $j$-th layer with refractive index of $n_j$ and thickness $d_j$: $\beta_j = \frac{2\pi}{\lambda} n_j d_j$.

The refractive index of $WO_3$ is obtained from its dielectric constant $n = \sqrt{\varepsilon(\omega)}$. In the data fitting, we used the Lorentz oscillator model, $\varepsilon(\omega) = \varepsilon(\infty) + \frac{\omega_p^2}{\omega_0^2 - i\gamma\omega - \omega^2}$; here, $\omega_p$ is the plasma frequency, $\gamma$ is the damping constant, and $\omega_0$ is the resonance frequency.[3] We use a high-frequency relative dielectric constant $\varepsilon(\infty) = 4.8$ for $WO_3$ thin films,[4,5] and the refractive indices $n = 1.43$ for the ionic liquid[6] and $n = 1.96$ for $YAlO_3$.[7]

The measured FTIR data are then fitted with $1 - \frac{|\mathcal{T}_4|^2}{|\mathcal{T}_4(\omega_p \to 0)|^2}$. Three independent parameters are to be determined: $\omega_0, \gamma$, and $\omega_p$. Using the least-squares method, these three parameters can be determined simultaneously for each measured FTIR curve. In Fig. S9 we show that as the charging process takes place, $\omega_0$ and $\gamma$ remain almost unchanged while $\omega_p$ increases significantly. Thus, we can take the average values of $\omega_0$ (0.43 eV) and $\gamma$ (0.77 eV) as fixed/known parameters and do the fitting again; $\omega_p$ is then determined and the free carrier concentration is derived from $\omega_p^2 = \frac{ne^2}{m\varepsilon_0}$.

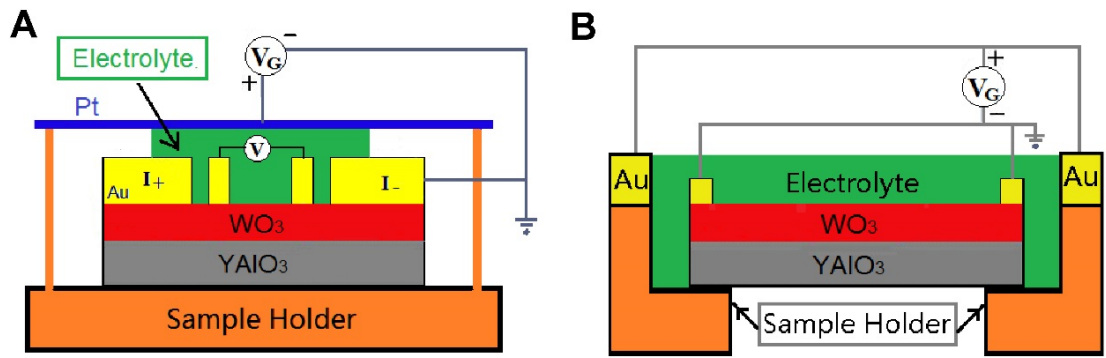

**Fig. S1. Sketches of devices for transport and optical measurements**. **a**, Top-gated devices for transport measurements, **b**, Side-gated devices for XRD and FTIR measurements.

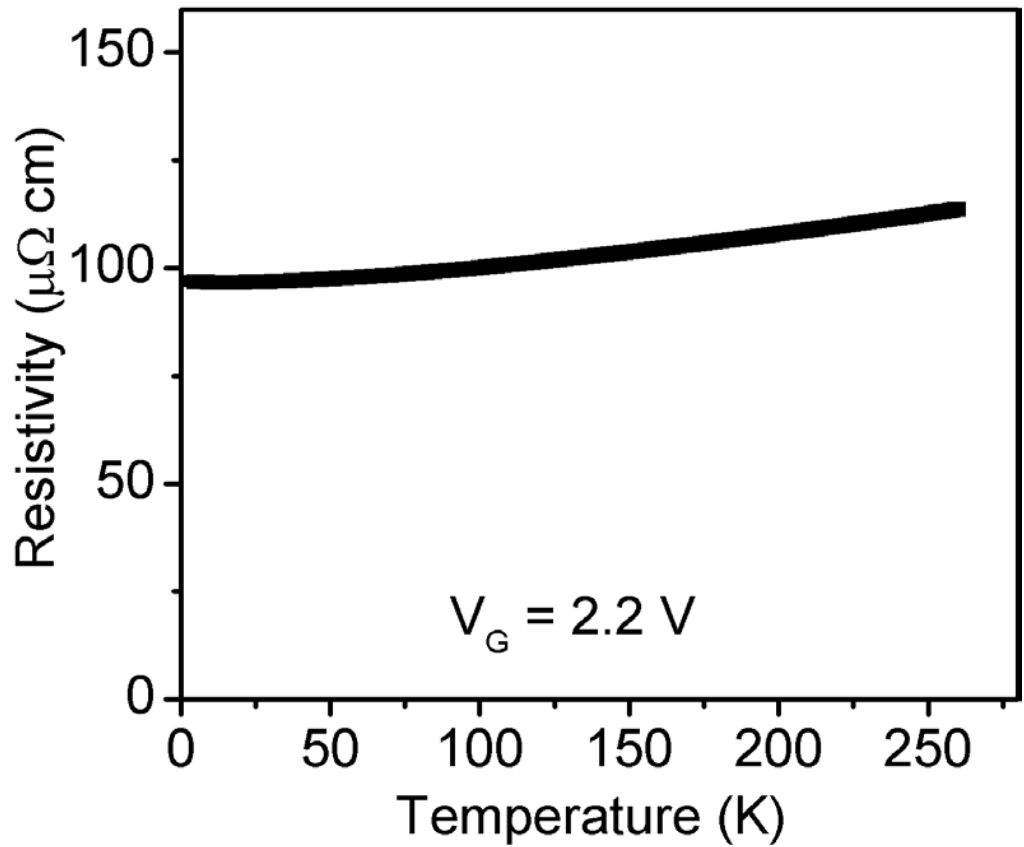

**Fig. S2. Resistivity measured down to $T$ = 300 mK for a 15 nm-thick $WO_3$ film charged to saturation.** No superconductivity can be seen. While at room temperature the charged $WO_3$ film is three times more conductive than an optimally-doped cuprate film, its temperature derivative of resistivity is much smaller and the residual resistivity much higher.

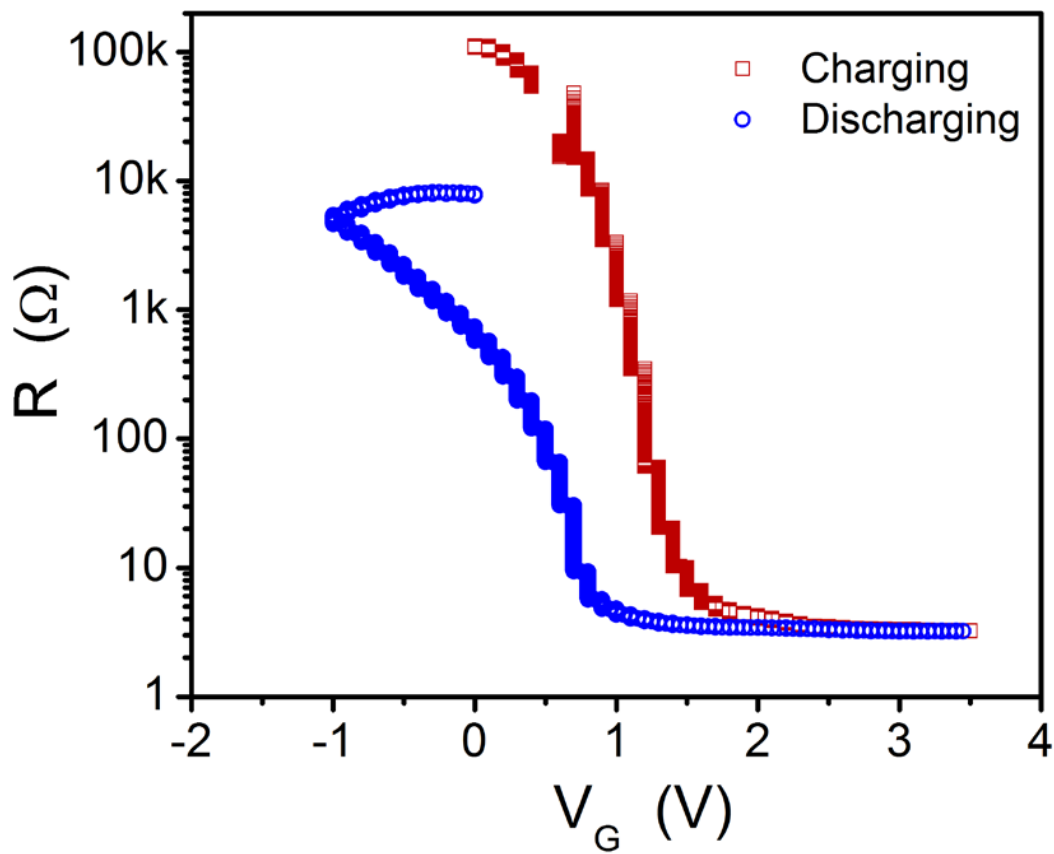

**Fig. S3. Resistance *vs.* gate voltage of a WO$_3$ film measured during charging and discharging process using pure polyethylene glycol (adding no salts whatsoever) as a gate dielectric.** The experiment was done at room temperature. The resistance drops by more than 4 orders of magnitude and saturates at $V_G$ = 2.2 V, the same as what we observe when using the ionic liquid DEME-TFSI or PEG-NaF as the gate electrolytes.

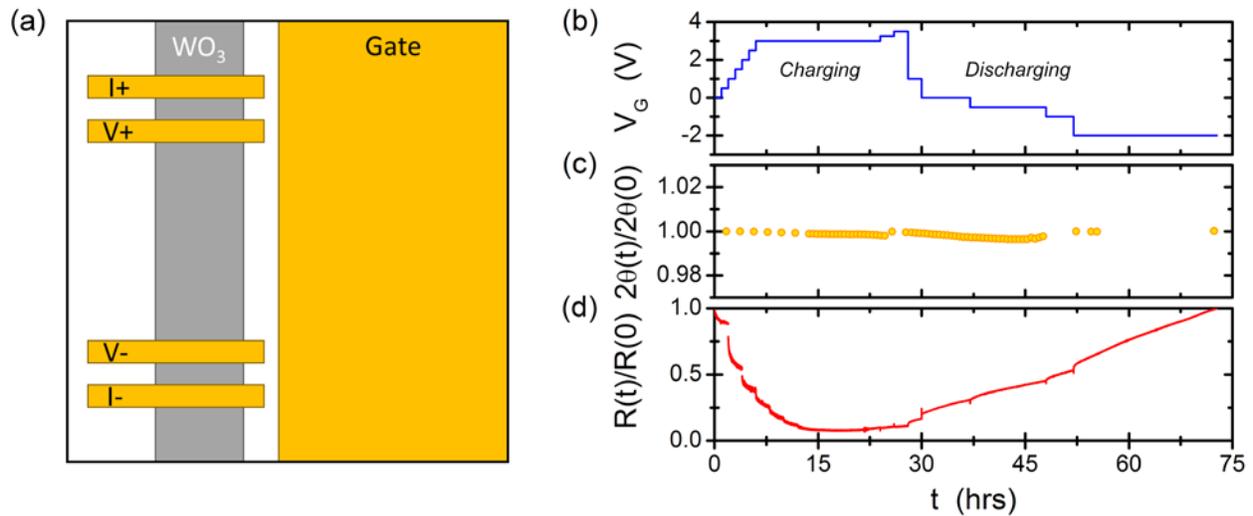

**Fig. S4. Combined *operando* XRD and transport measurements. a**, The sample that the *operando* XRD data shown in Fig. 5a was taken with was subsequently patterned into a 2 mm wide strip. Gold current and voltage contacts as well as a large coplanar gate were deposited and used for transport and gating experiments with PEG-NaF as the electrolyte. The voltage contacts had a center to center separation of 5 mm. **b**, The gate voltage was varied in time, charging first with gate voltages up to +3.5 V, and then discharging at -2 V. **c**, The $WO_3$ (002) peak position determined by XRD with the x-ray beam focused on the center of the strip was essentially constant the entire time. **d**, The resistance of the strip, however, was varied significantly throughout the measurement.

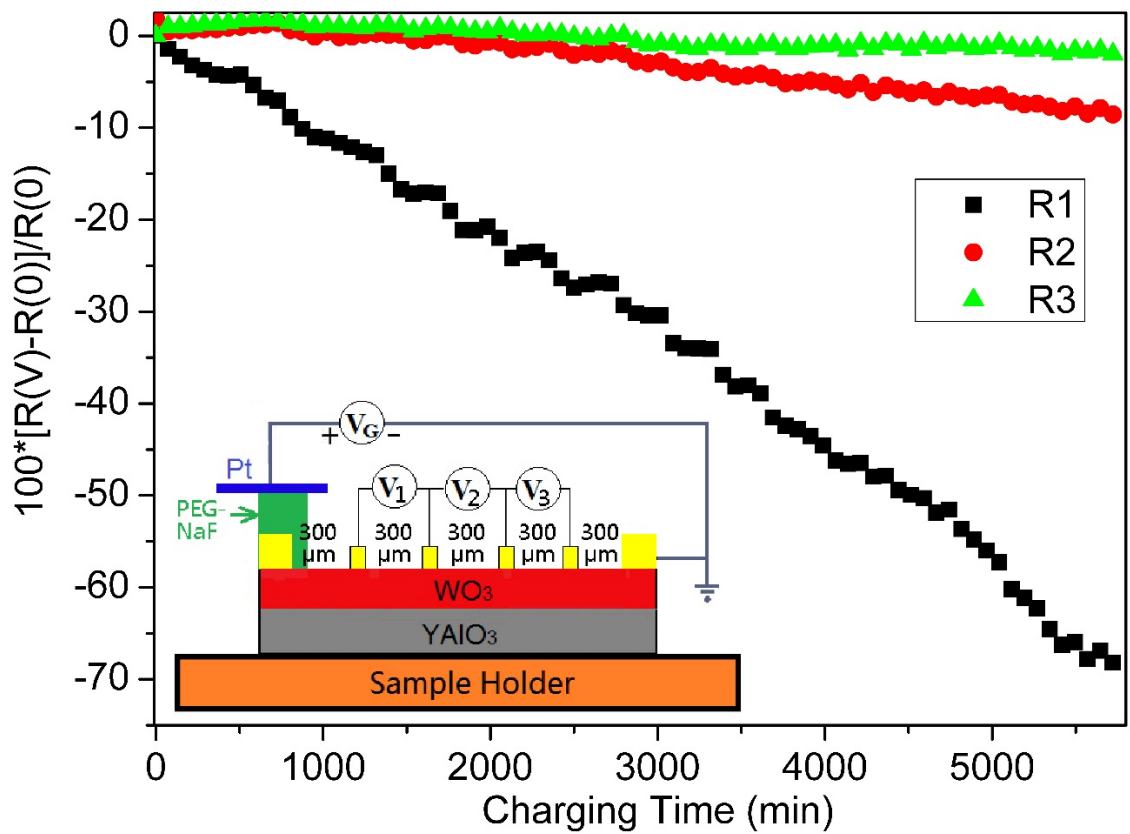

**Fig. S5. Resistances measured several hundred micrometers away from the charging area.** Note that in this configuration, the real gate voltage applied on the electrolyte (PEG-NaF) is lowered since the sample resistance is high. However, as long as $R_{electrolyte} > R_{sample}$, most of the gate voltage will be applied on the electrolyte. The gate current, which is in the range of several tens of nA after the first several minutes and decreases very slowly during the measurements, was used to calculate the resistance. The charging temperature was 305 K and the largest gate voltage we applied was 2.5 V. The observed large increase in conductivity in the portion of the film that was not covered with electrolyte indicates massive and easy electromigration. The effect only occurs for one polarity, with positive ions migrating outside of the gated area.

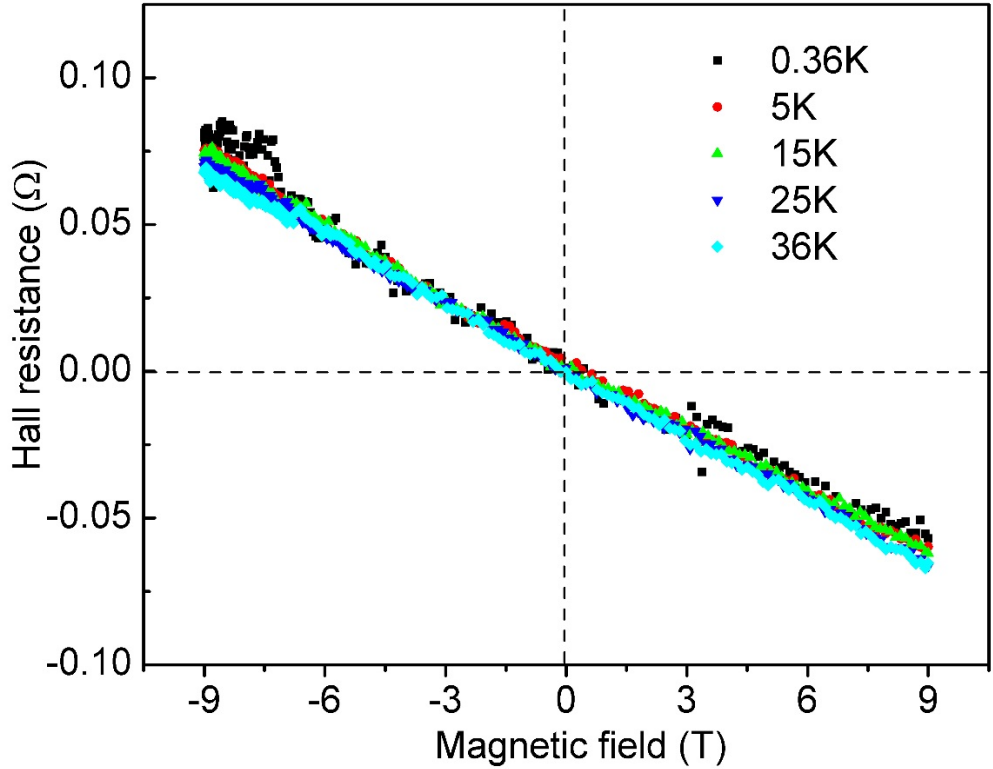

**Fig. S6 Hall resistance measured at different temperatures for a charged and saturated 15 nm WO₃ film.** The sign is negative, which indicates that the mobile charge carriers are electrons. The carrier density does not change with temperature. The calculated sheet carrier density reaches $8\times10^{16}$ cm$^{-2}$, which is much higher than what was reported in electrolyte gating studies so far.[8,9] Assuming that the entire film is uniformly doped, and using $R_\text{H} = 1/ne$, this would correspond to the 3D carrier density $n = 5.3\times10^{22}$ cm$^{-3}$, amounting to more than two electrons per one WO₃ unit cell. This seems unphysically high, and points instead to polaron formation.

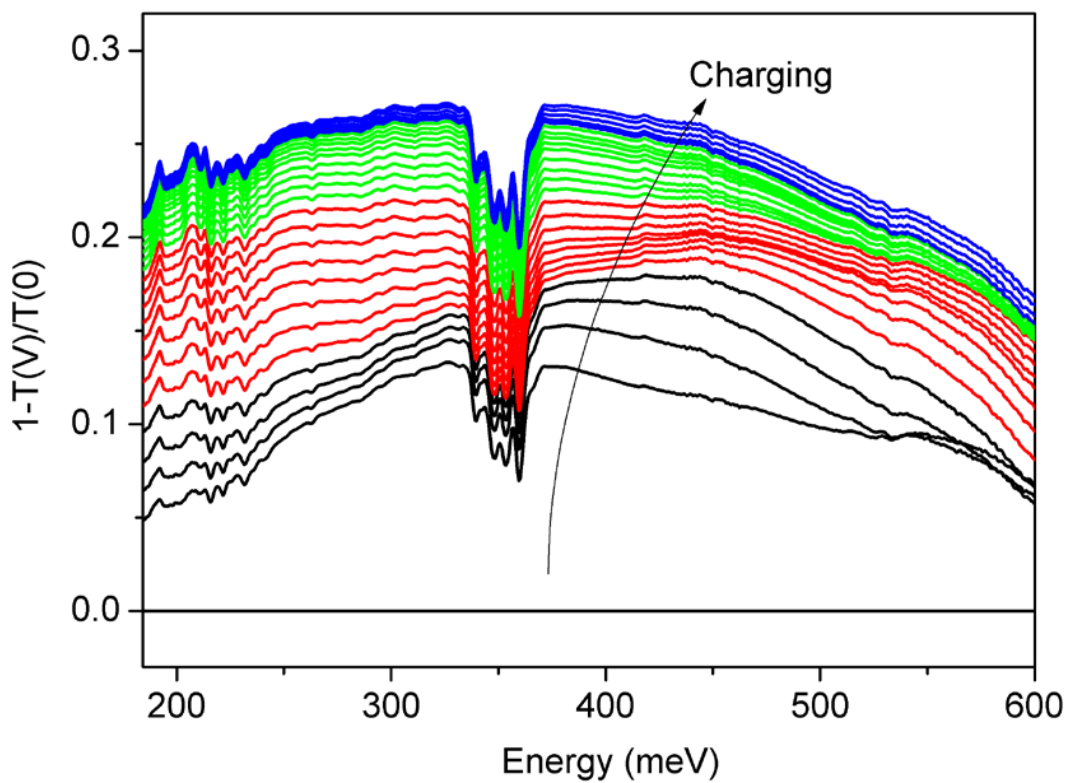

**Fig. S7. Infrared absorption spectra measured using the Fourier transform infrared spectroscopy during charging of a WO₃ film.** A thin Teflon foil held the ionic liquid in place. The electrolyte was DEME-TFSI and the charging was done at room temperature (294 K). The time interval between two successive curves was about 1 hour. We observed a similar peak when a glass coverslip was used instead of Teflon.

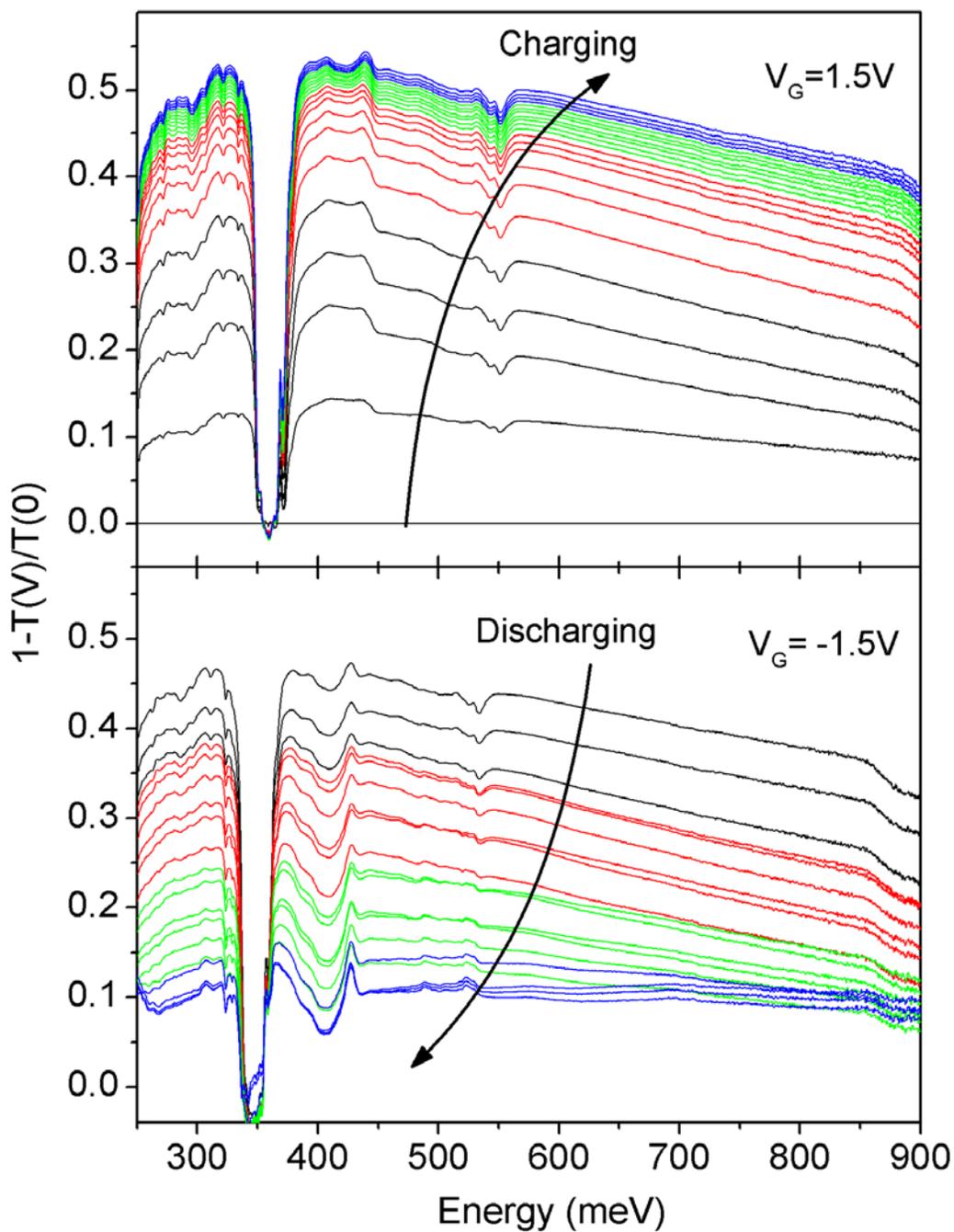

**Fig. S8. FTIR absorption spectrum measured during charging (upper panel) and discharging (bottom panel) processes.** The electrolyte was DEME-TFSI and the charging/discharging was done at room temperature (294 K). The time interval between two successive curves was about 1 hour. The first discharging curve was measured after we removed the positive gate voltage and applied the negative gate voltage immediately. Apparently, the charging process is reversible from the optical point of view.

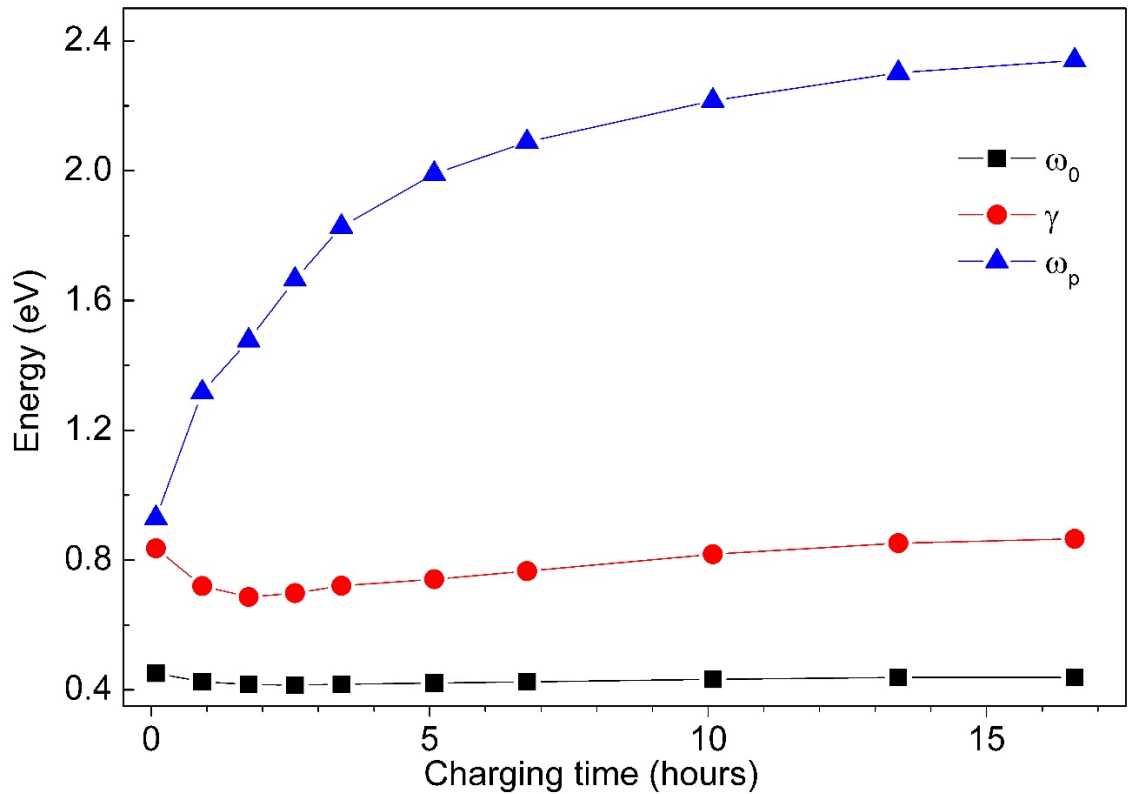

**Fig. S9. The FTIR data were fit by a model with three independent parameters: $\omega_0$, $\gamma$, and $\omega_p$, determined simultaneously using the least-squares method.** The resulting parameter values are plotted here as functions of the charging time.